\documentclass{egpubl}
\usepackage{vcbm2025s}
\WsShortPaper 

\usepackage[T1]{fontenc}
\usepackage{dfadobe}

\usepackage{cite}
\BibtexOrBiblatex
\electronicVersion
\PrintedOrElectronic
\ifpdf \usepackage[pdftex]{graphicx} \pdfcompresslevel=9
\else \usepackage[dvips]{graphicx} \fi
\usepackage{egweblnk}

\title[\toolname]%
      {\toolname: A Simple and Robust Detail-On-Demand Dashboard for Patient Data\vspace*{-0.8cm}}
\author[L. Schilcher, P. Waldert et al.]{
  \parbox{\textwidth}{
    \centering
    L. Schilcher$^{1}$\thanks{The first two authors contributed equally to this work.}\orcid{0009-0007-1051-2877},
    P. Waldert$^{1}$\footnotemark[1]\thanks{Corresponding Author. Email: peter.waldert@tugraz.at.}\orcid{0009-0004-8459-7381},
    B. Kantz$^{1}$\orcid{0000-0003-3294-8421}
    and T. Schreck$^{1}$\orcid{0000-0003-0778-8665} 
  } \\
  \parbox{\textwidth}{
    \centering
    $^1$ Institute of Visual Computing, Graz University of Technology
  }
}

\ccsdesc[500]{Human-centered computing~Visual analytics}
\ccsdesc[300]{Applied computing~Health informatics}

\usepackage{subfig}
\usepackage[nameinlink]{cleveref}
\usepackage{xcolor}
\usepackage{xspace}

\definecolor{panelonecolor}{HTML}{7593e6}
\definecolor{paneltwocolor}{HTML}{9ee5a1}
\definecolor{panelthreecolor}{HTML}{f3db70}
\newcommand{\toolname}{Clusters in Focus\xspace}
\newcommand{\PanelOne}{\textcolor{panelonecolor!80!black}{Panel 1}\xspace}
\newcommand{\PanelTwo}{\textcolor{paneltwocolor!80!black}{Panel 2}\xspace}
\newcommand{\PanelThree}{\textcolor{panelthreecolor!80!black}{Panel 3}\xspace}
\newcommand{\DataPanel}{\textcolor{panelonecolor!80!black}{Data Panel}\xspace}
\newcommand{\SelectionPanel}{\textcolor{paneltwocolor!80!black}{Selection Panel}\xspace}
\newcommand{\ClusterSimilarityPanel}{\textcolor{panelthreecolor!80!black}{Cluster Similarity Panel}\xspace}
\newcommand{\repourl}{\href{https://github.com/hereditary-eu/ClustersInFocus}{github.com/hereditary-eu/ClustersInFocus}\xspace}

\begin{document}
  \teaser{
    \vspace*{-1.0cm}
    \centering
    \includegraphics[width=0.933\linewidth]{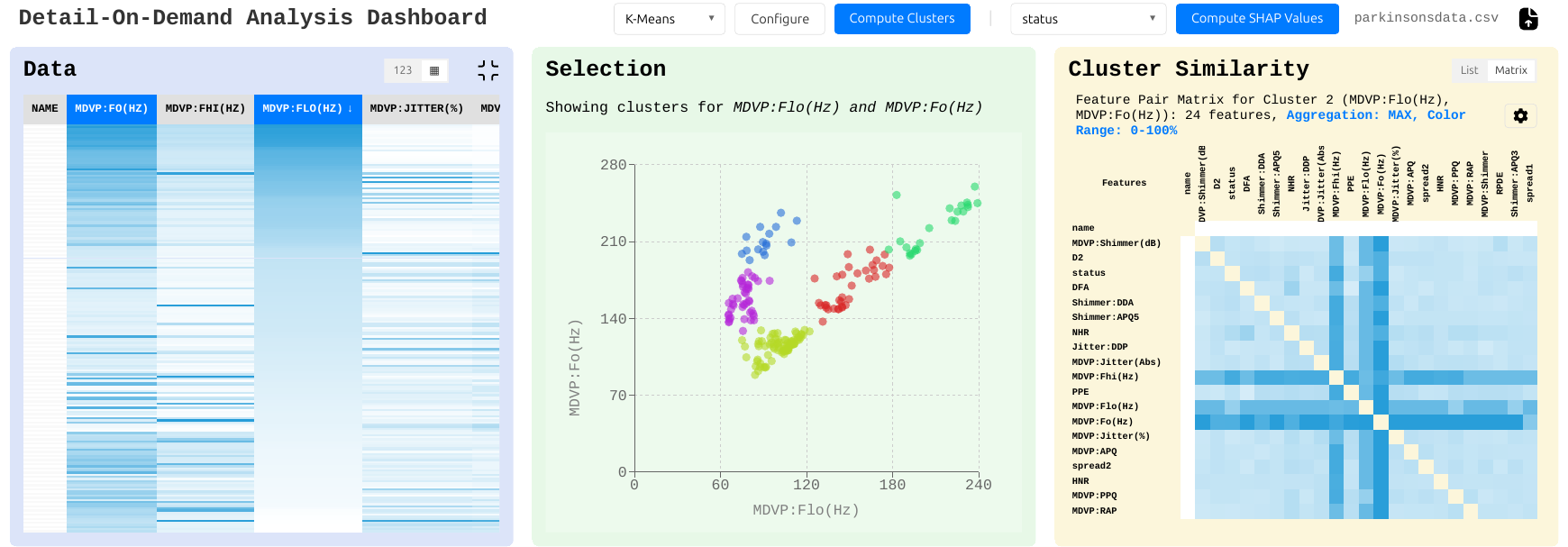}
    \vspace*{-1mm}
    \caption{Overview of \toolname: \DataPanel (left), \SelectionPanel (center), and \ClusterSimilarityPanel (right).}
    \label{fig:teaser}
  }

  \maketitle
  \begin{abstract}
    Exploring tabular datasets to understand how different feature pairs partition data into meaningful cohorts is crucial in domains such as biomarker discovery, yet comparing clusters across multiple feature pair projections is challenging. We introduce \textit{\toolname}, an interactive visual analytics dashboard designed to address this gap. \toolname employs a three-panel coordinated view: a \textbf{\DataPanel} offers multiple perspectives (tabular, heatmap, condensed with histograms / SHAP values) for initial data exploration; a \textbf{\SelectionPanel} displays the 2D clustering (K-Means/DBSCAN) for a user-selected feature pair; and a novel \textbf{\ClusterSimilarityPanel} featuring two switchable views for comparing clusters. A ranked list enables the identification of top-matching feature pairs, while an interactive similarity matrix with reordering capabilities allows for the discovery of global structural patterns and groups of related features. This dual-view design supports both focused querying and broad visual exploration. A use case on a Parkinson's disease speech dataset demonstrates the tool's effectiveness in revealing relationships between different feature pairs characterizing the same patient subgroup.
    \printccsdesc
  \end{abstract}

  \section{Introduction}
  In many applications, data items need to be understood in terms of groups of similar items. When data is described by many attributes (or features), the question arises in which features the grouping is present. While often, \emph{all} features are used to compare data simultaneously, it is also interesting to look for \emph{subsets} of features in which data can be grouped. This can  for example be relevant in multidimensional biomedical datasets, where researchers want to discover potential biomarkers (objective measures indicating a particular biological state or condition), and understand disease subtypes described in different data features~\cite{ritchie_methods_2015}. A key challenge in this process involves understanding how different combinations of features, especially pairs, relate to underlying patterns or partition the data into meaningful cohorts or clusters. Even when established biomarker pairs are known, discovering alternative or related feature pairs that characterize a similar cohort of subjects (e.g., patients) can lead to novel hypotheses or reveal previously overlooked correlations. Traditional exploratory data analysis often relies on examining feature pairs individually, for example through scatterplots, as these provide directly interpretable views based on original data dimensions. A systematic comparison of all possible subgroups or clusters becomes time-intensive and quickly infeasible as the number of features grows (and the number of pairs, accordingly). Furthermore, clustering based on the full feature set or using dimensionality reduction techniques may obscure patterns prominent only within specific low-dimensional subspaces, introduce hard-to-explain abstract features, and is sensitive to the curse of dimensionality~\cite{bellman1966curse}. \textit{Therefore, our approach specifically focuses on analyzing feature pairs}, which maintains interpretability by operating directly on original feature combinations.
  Existing visual analytics systems, such as Caleydo~\cite{2010-caleydo} and LineUp~\cite{2013-lineup}, provide coordinated views and interactive ranking mechanisms for multivariate data exploration. However, they do not specifically address the systematic comparison of cluster compositions generated from different feature subspaces, especially from feature pair projections. Tools like VICTOR~\cite{karatzas_victor_2021}, Clustrophile~\cite{demiralp2017clustrophiletoolvisualclustering}, and Clustrophile~2~\cite{cavallo2018clustrophile2} primarily support comparing different clustering algorithms or tuning parameters on a fixed feature space.

  Our design is motivated by a core analytical task: once users identify a meaningful cohort using one feature pair, they need a systematic way to discover other pairs that characterize a similar subgroup. This process is essential for validating findings and comparing related biomarkers within an interpretable, low-dimensional context.
  To address this gap, we present \toolname, an interactive visual analytics dashboard designed for the exploration of tabular data, aimed at identifying related feature pairs by comparing the similarity of their induced clusters. \toolname implements a coordinated three-panel workflow: \begin{enumerate}
    \item The \textbf{\DataPanel} provides multiple views (tabular, heatmap, condensed with histograms and optional SHAP-based feature importance~\cite{2017-shap}) for initial data overview and feature selection.
    \item The \textbf{\SelectionPanel} displays the 2D clustering (using K-Means or DBSCAN) of data points based on two user-selected features.
    \item The \textbf{\ClusterSimilarityPanel}, our core contribution, activates when a user selects a data point (and thus its cluster) in the \SelectionPanel. It quantitatively compares this source cluster against clusters from all other feature pairs using the Jaccard Index.  The results are presented in two complementary views: a \textbf{ranked list} for targeted identification of top-matching pairs, and a \textbf{reorderable similarity matrix} to reveal global patterns and relationships between feature groups.
  \end{enumerate}
  This workflow allows domain experts, such as biomedical researchers, to start with a known feature pair of interest, identify a relevant cluster, and efficiently discover other feature pairs that might serve as alternative or complementary biomarkers characterizing a similar subgroup.

  \section{Related Work}
  Our work intersects with several areas in visual analytics, primarily with interactive clustering analysis and visual cluster comparison.
  Due to the vast number of related works in visual cluster analysis, we can only refer to a limited selection; it is important to note that this field has been extensively researched for many years, and our references represent just a small sample of the existing literature.

  \textbf{Interactive Clustering Tools:}
  A rich body of work exists on systems that allow users to guide the clustering process. These tools support this goal through various means, such as iterative parameter tuning (Clustrophile 2~\cite{cavallo2018clustrophile2}), direct manipulation of cluster labels and their corresponding 2D projections (Cluster Sculptor~\cite{clustersculptor}), or visual supervision of results via quality metrics (Clustervision~\cite{clustervision}). While these systems excel at helping users optimize or validate a clustering configuration within a single, fixed feature space, \toolname's primary contribution is different. Our focus is on exploring the feature space itself by systematically comparing the resulting cluster structures across many different feature pair subspaces.

  \textbf{Visual Cluster Comparison Tools:}
  Tools like VICTOR~\cite{karatzas_victor_2021} enable users to compare pre-computed clustering partitions generated by different algorithms or parameters on the same dataset. VICTOR uses metrics like the Jaccard Index and Adjusted Rand Index, visualized through methods like heatmaps or Sankey diagrams, to assess the overall similarity between different partitioning strategies. While VICTOR excels at comparing partitions based on the same feature set, it does not directly address clusters formed from different feature combinations. \toolname fills this gap by systematically clustering across all feature pairs and providing ranked similarity comparisons centered on a user-selected cluster, thus comparing the \textit{results} of projecting data onto different feature pairs.

  \textbf{Visual Analytics for Multivariate Data:}
  Existing visual analytics systems for multivariate data form the basis for our design. Caleydo~\cite{2010-caleydo} integrates tabular data (e.g., gene expression) with contextual information (e.g., biological pathways) using multiple coordinated views, with a focus on biological interpretation. While \toolname also utilizes coordinated views, it specifically supports workflows for comparing cluster membership similarity across different feature pair views within a single table, without integrating external context. LineUp~\cite{2013-lineup} enables interactive ranking of items based on user-defined combinations of attributes. \toolname's \PanelThree presents a ranked list that ranks (feature pair, cluster ID) tuples by their Jaccard similarity to a reference cluster to identify feature pairs yielding similar data cohorts rather than ranking individual items by attribute scores.

  \textbf{Subspace Analysis and Visualization:}
  Our work relates to the visual analysis of feature subspaces. Several systems support this exploration with different goals. For instance, some tools visually compare 2D subspaces by ranking them according to various quality metrics~\cite{DBLP:conf/ieeevast/TatuMFBSSK12}. The Patterntrails approach~\cite{DBLP:conf/ieeevast/JackleHBKS17} traces data points across an ordered sequence of subspaces to reveal their behavior. A related approach is "SmartStripes" by May et al.~\cite{reviewer-smart-stripes-ref}, which guides feature subset selection for predictive modeling. Their system decomposes global statistical quality measures across data partitions, visualizing local importance for specific cohorts and interactively steering a selection algorithm.

  Based on the indicated problems, \toolname differentiates itself from these methods by focusing on the direct comparison of data cohorts. Instead of ranking subspaces by abstract statistical metrics for model building~\cite{DBLP:conf/ieeevast/TatuMFBSSK12, reviewer-smart-stripes-ref} or tracing individual points~\cite{DBLP:conf/ieeevast/JackleHBKS17}, our approach is centered on a user-selected cluster. It then systematically discovers other feature pairs which partition the data similarly, directly addressing the task of finding related features that characterize the same underlying group.

  \section{The \toolname -Tool}
  \toolname provides an interactive web-based environment for analyzing feature pair relationships in tabular data via clustering. Its three coordinated panels support a structured workflow from initial overview to detailed cluster comparison (\Cref{fig:teaser}).

  \textbf{Implementation Architecture:}
  \toolname follows a client-server architecture to outsource computational demands while keeping interactive performance. The frontend is built with React and TypeScript and provides a responsive user interface running entirely in the user's web browser. A lightweight Python backend using FastAPI handles computational tasks.
  With this separation, the frontend focuses exclusively on visualization and user interaction, while the backend manages analytical processing and result caching.
  The full source code is available at \repourl and the application has been dockerized.
  To build and start the application, run \texttt{docker compose build} and \texttt{docker compose up}.

  \subsection{\PanelOne: Data \& Feature Exploration}
  Once a CSV dataset has been uploaded via the upload control in the application's header, \PanelOne provides the primary interface for feature exploration and selection. This panel offers three distinct views, switchable via icons:

  \begin{figure}[t]
    \vspace*{-3mm}
    \centering
    \subfloat[\textbf{Heatmap View}: Condensed overview of normalized feature correlations in \PanelOne, similarly arranged as in \textit{LineUp} \cite{2013-lineup}.]{
      \includegraphics[width=0.44\linewidth, clip]{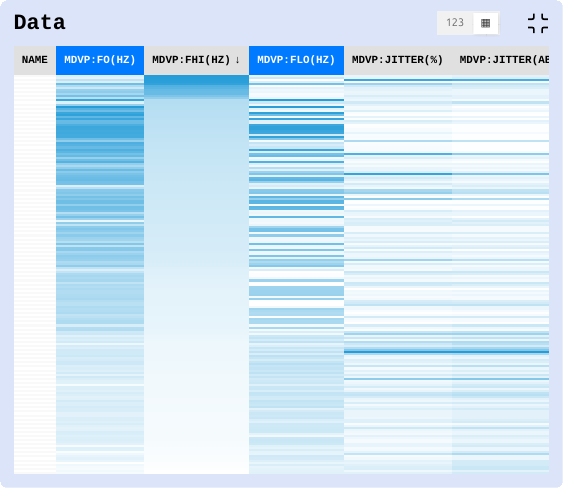}
      \label{fig:panel1_heatmap}
    }
    \hfill
    \subfloat[\textbf{Condensed View}: Miniature histograms per feature, with an optional SHAP-based feature ranking~\cite{2017-shap} in \PanelOne.]{
      \includegraphics[width=0.44\linewidth, trim={0 6.2cm 0 0}, clip]{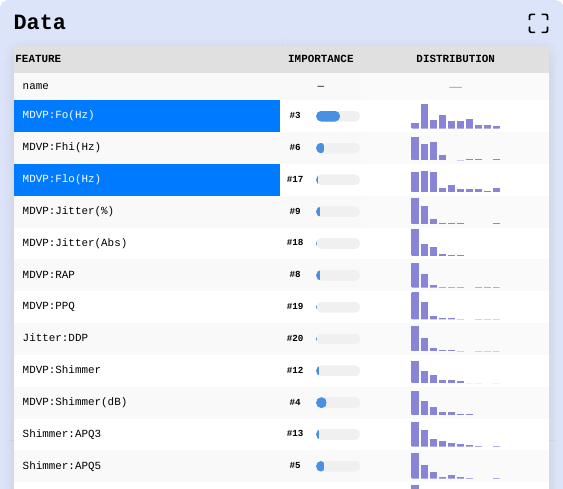}
      \label{fig:panel1_condensed}
    }
    \vspace*{-2mm}
    \caption{\PanelOne views in \toolname: (a) Heatmap and (b) Condensed feature summary.}
    \label{fig:panel1_views}
    \vspace*{-5mm}
  \end{figure}

  \begin{itemize}
    \item \textbf{Tabular View:} Presents the data in a standard spreadsheet-like table. Users can sort the table rows based on the values in any column and hide columns to focus the view.
    \item \textbf{Heatmap View:} Provides a condensed overview of numerical features. Each cell's value is mapped to a color gradient (normalized within its column), and rows are shrunk vertically to maximize the number of visible items (\Cref{fig:panel1_heatmap}). Hovering over a row expands it to reveal the actual numerical values which allows quick inspection within the overview context. Similar to the Tabular View, columns can be sorted or hidden.
    \item \textbf{Condensed View:} Offers a feature-centric summary (\Cref{fig:panel1_condensed}). It lists all features, showing a miniature histogram of the respective distribution. Optionally, if SHAP values have been computed (by selecting a target variable and clicking "Compute" in the header), this view displays the rank of each feature's importance. Clicking on a mini-histogram opens a larger view with configurable bin count for detailed distribution analysis.
  \end{itemize}
  Regardless of the active view within \PanelOne, users select features for analysis in \PanelTwo by clicking directly on the respective column headers. \toolname enforces a limit of exactly two active features for the subsequent clustering visualization. If a user selects a third feature header, the chronologically first selected feature is automatically deselected to ensure that only the two most recently selected features are passed to the \SelectionPanel.

  \subsection{\PanelTwo: Focused 2D Clustering}
  If two features $(F_A, F_B)$ are selected in \PanelOne, the central \SelectionPanel displays the corresponding 2D scatterplot of the data points (\Cref{fig:teaser}, center panel). The points in this scatterplot are colored to reflect their cluster assignments. These assignments are derived from clustering results computed using the algorithm and parameters specified via controls in the application header. Specifically, the header controls allow the user to: \begin{itemize}
    \item Select the clustering algorithm (K-Means or DBSCAN) via a dropdown menu.
    \item Configure the algorithm's hyperparameters (e.g., 'k' for K-Means; 'eps' and 'min\_samples' for DBSCAN).
    \item Trigger the backend clustering (re-)computation (or re-computation if parameters change) for the currently selected feature pair $(F_A, F_B)$.
  \end{itemize}
  Clustering results are retrieved from the backend after computation, and points are colored according to their assigned cluster IDs in the \PanelTwo scatterplot.
  In principle, the number of clusters (and therefore, colors) is not limited from a technical viewpoint, but of course after a certain number ($\sim 10$) they become hard to distinguish visually.
  The primary interaction \textit{within} \PanelTwo itself is \textbf{selecting a data point.} Clicking on any point selects the cluster it belongs to and populates \PanelThree with the cluster similarity analysis.

  \subsection{\PanelThree: Cross-Pair Cluster Similarity}
  \begin{figure}[t]
    \centering
    \includegraphics[width=\linewidth, trim={0 4.9cm 0 0}, clip]{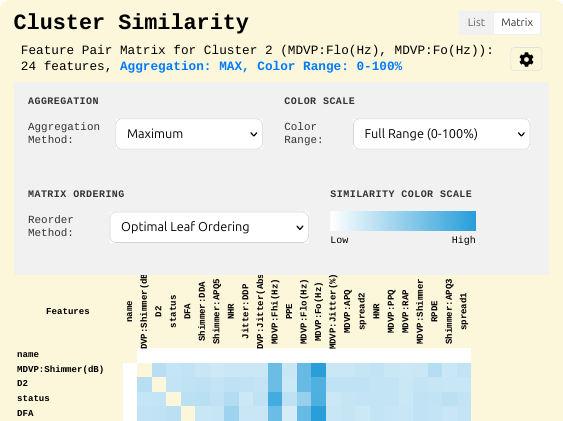}
    [cut off; a full view is available in \Cref{fig:teaser}.]
    \vspace*{-2mm}
    \caption{
      The \ClusterSimilarityPanel in Matrix View configuration. The panel provides two switchable views (top right): a List View for targeted ranking and this Matrix View for global pattern discovery.
      Each cell's color represents the aggregated Jaccard similarity, with the aggregation method (Maximum) and matrix ordering (Optimal Leaf Ordering) selected by the user. The reordering algorithm groups features with similar profiles, revealing a distinct high-similarity block around the source features and exposing, through the bright horizontal and vertical lines, their strong relationship with other feature groups related to vocal stability (e.g., \texttt{HNR}, \texttt{DFA}) and jitter (\texttt{Jitter:DDP}).
    }
    \label{fig:panel3_matrix}
    \vspace*{-5mm}
  \end{figure}

  This rightmost panel is the core analytical component of \toolname, activating upon selection of a data point (and thus its cluster) in \PanelTwo. Let the features selected in \PanelOne be $F_A$ and $F_B$, and let the user-selected cluster be $C_{AB}$ (representing a specific set of data points). \PanelThree then performs a systematic comparative analysis before presenting the results in one of two switchable views.
  The underlying analysis proceeds as follows:
  \begin{itemize}
    \item For \textit{every other unique pair} $(F_X, F_Y)$ in the dataset, the tool retrieves or computes the clustering result obtained using the same algorithm and parameters specified in the header. This yields a set of clusters $\{C_{XY:1}, C_{XY:2}, \cdots, C_{XY:k}\}$ for each pair $(F_X, F_Y)$.
    \item The Jaccard similarity $J(C_{AB}, C_{XY:k})$ is calculated between the initially selected cluster $C_{AB}$ and \textbf{every individual cluster} $ C_{XY:k}$ from \textbf{every other feature pair} $(F_X, F_Y)$. The Jaccard Index $J \in [0, 1]$ measures the overlap in data point membership $$J = \frac{\left| C_{AB} \cap C_{XY:k} \right|}{\left| C_{AB} \cup C_{XY:k} \right|}\,.$$
  \end{itemize}
  These calculated similarity scores form the basis for the two complementary views offered by the panel:
  \begin{itemize}
    \item \textbf{List View}: This view directly utilizes the individual similarity scores. It presents a table that ranks all (Feature Pair, Cluster ID) combinations based on their calculated Jaccard similarity with the source cluster $C_{AB}$, from highest to lowest. This view is ideal for focused queries to quickly identify the single best-matching clusters from other feature projections.
    \item \textbf{Matrix View}: This view (\PanelThree) provides a global, structural overview. Instead of listing individual cluster scores, it aggregates them. For each feature pair $(F_X, F_Y)$, it combines the set of similarity scores $\{J(C_{AB}, C_{XY:1}),\, J(C_{AB}, C_{XY:2}), ...\}$ into a single value using a user-selected aggregation method (e.g., Maximum, Average). The result is a feature-by-feature heatmap where cell $(F_X, F_Y)$ is colored by this aggregated score. With interactive matrix reordering (e.g., Optimal Leaf Ordering), this view aims at revealing high-level patterns, such as blocks of related features that consistently define similar cohorts.
  \end{itemize}
  By offering both views, the system supports both focused, query-driven analysis and broader, discovery-oriented exploration of the dataset's cluster relationships.

  \textbf{Workflow Summary:}
  \hspace{1mm}\textbf{1.} Upload a dataset (CSV).
  \hspace{1mm}\textbf{2.} Explore features in \PanelOne (Table, Heatmap, Condensed views) to assess distributions and SHAP-ranked importance (optional).
  \hspace{1mm}\textbf{3.} Select two features $(F_A, F_B)$.
  \hspace{1mm}\textbf{4.} Configure parameters and compute clusters in \PanelTwo.
  \hspace{1mm}\textbf{5.} Select a cluster $C_{AB}$ by clicking a data point.
  \hspace{1mm}\textbf{6.} Analyze the results in \textbf{\PanelThree}. Use the \textbf{List View} for a ranked overview of top matches, or switch to the \textbf{Matrix View} and apply reordering to discover higher-level patterns and relationships between feature groups.

  \section{Use Case: Parkinson's Disease Speech Biomarkers}
  We demonstrate \toolname using the Parkinson's Disease speech dataset from the UCI Machine Learning Repository~\cite{parkinsons_174}, containing voice measurements and a 'status' attribute (healthy/PD). The goal is to identify speech patterns that differentiate the groups.

  The analyst loads the dataset and selects \texttt{MDVP:Flo(Hz)} (minimum fundamental frequency) and \texttt{MDVP:Fo(Hz)} (average fundamental frequency) in \PanelOne. Using the header controls, they choose K-Means clustering, set $k=5$, and trigger the cluster computation. \PanelTwo then displays the 2D scatterplot, coloring points based on their cluster assignments (\Cref{fig:teaser}). The analyst observes the clusters and decides to investigate 'Cluster 2' further by clicking on a point within it, hypothesizing that it represents a specific patient subgroup.

  \PanelThree updates to display the similarity analysis for 'Cluster 2'. The analyst first inspects the \textbf{List View}, which confirms that other feature pairs, especially those involving vocal perturbation, can also identify highly similar cohorts. For a more comprehensive overview, they switch to the \textbf{Matrix View} and apply \textit{Optimal Leaf Ordering} (cf. \Cref{fig:teaser}). The reordering reveals a distinct block of high-similarity cells, grouping the fundamental frequency measures with features related to non-linear dynamics (\texttt{DFA}, \texttt{PPE}, \texttt{spread1}) and vocal stability (\texttt{HNR}). This indicates that 'Cluster 2' is not an artifact of a single feature pair, but a stable cohort identifiable across a family of different acoustic measurements.

  \textbf{Interpretation:}
  The analysis reveals that the cohort identified by 'Cluster 2', based purely on fundamental frequency measures, is consistently captured by a broader family of acoustic features. The high Jaccard similarities across this block, made evident by the Matrix View, suggest that 'Cluster 2' represents a robust subgroup characterized by a distinct and multi-faceted acoustic profile. The substantial overlap observed between this cluster and the cohort defined by the clinical \texttt{status} label supports the clinical relevance of this discovered subgroup. This workflow demonstrates how \toolname facilitates the validation of an initial observation, by highlighting that an identified pattern extends across multiple features and corresponds to a clinically significant population.

  \section{Discussion}
  \toolname is an easy-to-use dashboard for analysis of clusters in data. Its focus is on the comparison of clusters in multiple 2-dimensional subspaces. Thereby, the results of \toolname are easily interpretable and can be visually comprehended, as 2-dimensional subspaces can be directly visualized in scatter plots. A main application, and motivation for its development, are analysis tasks in biomedical applications. For example, grouping patients, or medical compounds into groups according to interpretable features, is often important for making conclusions. The results can be presented in a straightforward way. Future work should explore the benefits and limitations of \toolname, considering visual analytics approaches for subspace analysis in high-dimensional subspaces exist. Note that the latter often have to employ feature selection or dimensionality reduction to be interpretable, and may hence incur a presentational bias.

  \toolname allows comparing a selected cluster across different subspaces by offering two views. Alongside a \textbf{tabular list view}, it provides an \textbf{interactive heatmap matrix} to overview the Jaccard index across all pairwise dimension combinations. This matrix incorporates sorting capabilities to help identify block patterns, allowing for conclusions on influential and similar features. This approach provides a high-level view related to the comparison of cluster properties across subspaces, as discussed in \cite{DBLP:conf/ieeevast/JackleHBKS17}, by showing the stability of a cohort's composition as the underlying feature pair changes.

  While effective for pairwise exploration, the current approach is limited by the combinatorial growth in feature pairs for very high-dimensional datasets and its focus on 2D projections. Furthermore, practical tests indicate a scalability limit, as the backend's in-memory data processing and the frontend's direct rendering of all data points (which can be easily addressed in future versions of the tool) can lead to unresponsiveness with datasets exceeding several megabytes. Future work will explore methods for handling or selecting higher-dimensional feature combinations, likely guided by either dimensionality reduction techniques or feature importance measures. We also plan to extend the cluster comparison functionality by integrating other similarity metrics (e.g., Adjusted Rand Index) and exploring asymmetric or subset-aware measures to better capture relationships between clusters of differing sizes. To improve usability and lower the entry barrier, we include a pre-loaded default dataset. While the tool's three-panel design provides an implicit workflow following Shneiderman's \emph{Overview first, zoom and filter, details-on-demand} mantra \cite{Shneiderman1996}, we acknowledge that more explicit user guidance is an important next step.

  As demonstrated in the use case discussion, we have indication \toolname is able to provide interesting analysis across feature spaces. Future work should refine the requirement analysis, and apply and evaluate the approach in different domain settings.

  \section{Conclusion}
  We presented \toolname, an interactive visual analytics tool for exploring tabular data through cluster similarity comparisons across different 2D feature pair projections. By integrating data overview techniques, interactive 2D clustering, and a cross-pair comparison panel ranked by Jaccard similarity, \toolname enables users to efficiently identify feature pairs that capture similar data cohorts. Our approach provides a simple, fast exploration. Future work will expand and validate the approach.

  \section*{Acknowledgements}
  This work was supported by the HEREDITARY Project, as part of the European Union's Horizon Europe research and innovation programme under grant agreement No GA 101137074.
  Part of this work has already been outlined in a technical report, Deliverable 5.1 \cite{hereditary-deliverable-51}.

  \bibliographystyle{eg-alpha-doi}
  \bibliography{sources}
\end{document}